\documentclass[aps,letterpaper,twocolumn,prl,showpacs]{revtex4-1}
\usepackage{amssymb}
\usepackage{amsmath,bm}

\usepackage[pdftex]{graphicx,color}
\usepackage[english]{babel}
\usepackage{hyperref}

\begin{document}

\title{Harmonic and rational harmonic driving of microresonator soliton frequency combs}

\author{Yiqing Xu$^{1,2}$}
\author{Yi Lin$^{3}$}
\author{Alexander Nielsen$^{1,2}$}
\author{Ian Hendry$^{1,2}$}
\author{St\'ephane Coen$^{1,2}$}
\author{Miro Erkintalo$^{1,2}$}
\author{Huilin Ma$^{3}$}
\author{Stuart G. Murdoch$^{1,2}$}
\email{s.murdoch@auckland.ac.nz}

\affiliation{$^1$The Dodd-Walls Centre for Photonic and Quantum Technologies, Auckland 1010, New Zealand}
\affiliation{$^2$Department of Physics, University of Auckland, Auckland 1010, New Zealand}
\affiliation{$^3$School of Aeronautics and Astronautics, Zhejiang University, 310027, Hangzhou, China}
\affiliation{e-mail: mahl@zju.edu.cn}

\begin{abstract}
With demonstrated applications ranging from metrology to telecommunications, soliton microresonator frequency combs have emerged over the past decade as a remarkable new technology. However, standard implementations only allow for the generation of combs whose repetition rate is tied close to the fundamental resonator free-spectral range (FSR), offering little or no dynamic control over the comb line spacing. Here we propose and experimentally demonstrate harmonic and rational harmonic driving as novel techniques that allow for the robust generation of soliton frequency combs with discretely adjustable frequency spacing. By driving an integrated Kerr microresonator with a periodic train of picosecond pulses whose repetition rate is set close to an integer harmonic of the 3.23~GHz cavity FSR, we deterministically generate soliton frequency combs with frequency spacings discretely adjustable between 3.23~GHz and 19.38~GHz. More remarkably, we also demonstrate that driving the resonator at rational fractions of the FSR allows for the generation of combs whose frequency spacing corresponds to an integer harmonic of the pump repetition rate. By measuring the combs' radio-frequency spectrum, we confirm operation in the low-noise soliton regime with no supermode noise. The novel techniques demonstrated in our work provide new degrees of freedom for the design of synchronously pumped soliton frequency combs.

\end{abstract}

\maketitle

\section{Introduction}

\noindent Ever since their first demonstration~\cite{DelHaye2007}, microresonator optical frequency combs (``microcombs'')~\cite{kippenberg_microresonator-based_2011, pasquazi_micro-combs:_2018, Kippenberg2018, gaeta_photonic-chip-based_2019} have attracted significant interest due to their wide range of potential applications in fields as diverse as metrology~\cite{suh_microresonator_2016, dutt_-chip_2018, Lamb2018}, optical frequency synthesis~\cite{Spencer2018}, imaging~\cite{bao_microresonator_2019}, distance measurements~\cite{Trocha2018, suh_soliton_2018}, and telecommunications~\cite{MarinPalomo2017}. Key to unleashing their potential has been the discovery that microresonators can support stable localized dissipative structures referred to as temporal cavity solitons (CS), or alternatively dissipative Kerr solitons~\cite{Herr2013}. These structures -- first observed in macroscopic fibre ring resonators~\cite{leo_temporal_2010} -- correspond to pulses of light that can circulate around the resonator without changes in their shape or energy~\cite{Coen2013,Coen2013_2}; in the spectral domain, they manifest themselves as coherent optical frequency combs~\cite{erkintalo_coherence_2014, yi_soliton_2015, brasch_photonic_2016, webb_experimental_2016, jang_synchronization_2018}.

Soliton microcombs are conventionally generated by driving a high-Q Kerr microresonator with a continuous wave (CW) laser~\cite{Herr2013, yi_soliton_2015, brasch_photonic_2016, webb_experimental_2016, jang_synchronization_2018}. Whilst appealing in its simplicity, this scheme suffers from several shortcomings. First, unless complicated control procedures are employed~\cite{guo_universal_2017, cole_kerr-microresonator_2018}, the number and relative positions of the excited solitons is essentially random. Second, the pump-to-comb power conversion efficiency is typically low, as only a small fraction of the pump field overlaps with the soliton (and hence transfers energy to the comb)~\cite{xue_microresonator_2017}. Third, excepting the formation of ``soliton crystals'' that hinge on difficult-to-control mode interactions~\cite{Cole2017, Karpov2019}, the comb repetition rate is restricted to a narrow range about the cavity free-spectral range (FSR), and can be challenging to lock to an external radio-frequency (RF) signal.

In a pioneering recent work, Obrzud et al. demonstrated that some of the limitations of CW-driven systems can be overcome by driving the resonator with a train of pulses synchronized to the round trip time of the cavity~\cite{Obrzud2017}. In this configuration, the soliton can be robustly ``trapped'' at a specific position within the intracavity pump pulse~\cite{hendry_spontaneous_2018}, thus locking the comb repetition rate to that of the driving pulse train whilst concomitantly suppressing its timing jitter~\cite{brasch_nonlinear_2019}. Although the initial experiments~\cite{Obrzud2017} used a pulse train whose repetition rate was matched to the fundamental cavity free-spectral range (FSR), more recent studies have extended the concept into the regime of sub-harmonic pumping, where the pump repetition rate is set close to one half of the cavity FSR~\cite{obrzud_microphotonic_2019, anderson2019photonic}. Yet, both of these schemes (synchronous and sub-harmonic) still share one of the limitations of CW-driven systems: they can only deliver microcombs whose repetition rate is close to the fundamental cavity FSR.

In this Article, we propose and experimentally demonstrate novel pulsed driving techniques that allow for the deterministic generation of soliton frequency combs with discretely controllable repetition rate. In particular, leveraging knowledge from the field of actively mode-locked lasers~\cite{Onodera1993,Jun1997,Chiming2000,Yoshida1996}, we show that driving a microresonator close to integer multiples or rational fractions of the cavity FSR enables microcombs whose repetition rate is close to an integer harmonic of the FSR. Our experiments are performed in a novel integrated silica waveguide resonator platform with an FSR of 3.23~GHz, and we generate low-noise soliton microcombs whose repetition rates range from the fundamental FSR (3.23~GHz) to its sixth harmonic (19.38~GHz). Significantly, by leveraging rational harmonic driving, we are also able to generate a microcomb at the second harmonic of the cavity FSR (6.46~GHz) despite using a driving pulse train at $2.15~\mathrm{GHz}\approx2/3~\mathrm{FSR}$. The new pump configurations demonstrated in our work expand the number of degrees of freedom available to microresonator designers, and could find useful application in areas that would benefit from adjustable comb spacings, such as reconfigurable optical communications networks.

\section{Concept}

\begin{figure}[t]	
		\includegraphics[width=\linewidth]{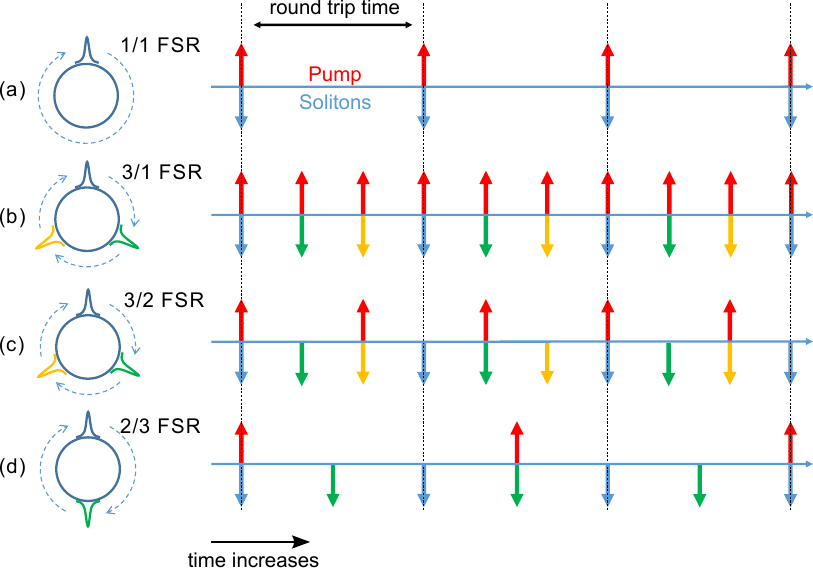}
		\caption{Visualization of harmonic and rational harmonic driving for different pump repetition rates: (a) $f_\mathrm{in}\approx \text{FSR}$, (b)~$f_\mathrm{in}\approx 3\,\text{FSR}$, (c) $f_\mathrm{in}\approx 3/2\,\text{FSR}$, (d) $f_\mathrm{in}\approx 2/3\,\text{FSR}$. The red arrows represent the input pump pulses, whilst the blue, green, and orange arrows represent the solitons.}
		\label{fig1}
\end{figure}

We consider a Kerr nonlinear microresonator driven with a train of pulses whose temporal duration is much shorter than the cavity round trip time (but much larger than the temporal width of the solitons). The repetition rate $f_\mathrm{in}$ of the input driving pulse train is set close to a rational fraction of the cavity FSR, i.e.,
\begin{equation}
f_\mathrm{in}\approx \frac{m}{n}\text{FSR},
\end{equation}
where $m$ and $n$ are integers with no common factors. In steady-state, the intracavity field is comprised of $m$ equally spaced pulses, each of which is repumped by the driving pulse train once every $n$ round trips [see Fig.~\ref{fig1}]. Each of the intracavity pulses can independently support a single soliton that is temporally locked at a set position along the pulse envelope~\cite{Obrzud2017, hendry_spontaneous_2018}; if the pump pulses are identical and the repetition rate $f_\mathrm{in}$ appropriately chosen (as elaborated below), each of the soliton trapping positions will be the same relative to its background pulse. In this case, the intracavity field will be composed of $m$ equally spaced solitons, yielding an output frequency comb with a repetition rate $f_\mathrm{out}$ that is exactly equal to the $n^\mathrm{th}$ harmonic of the input pulse train, i.e., $f_\mathrm{out} = nf_\mathrm{in}$, and approximately equal to the $m^\mathrm{th}$ harmonic of the cavity FSR. We must emphasize that an exact match between the repetition rate $f_\mathrm{in}$ and an integer faction of the cavity FSR is not required~\cite{Obrzud2017, hendry_spontaneous_2018}: robust trapping of CSs to the pump pulses can be achieved over a non-zero range of desynchronizations $f_\mathrm{in}-(m/n)\text{FSR}$. This trapping underpins the fact that the comb repetition rate will be exactly an integer harmonic of the pump repetition rate, and only approximately equal to an integer harmonic of the FSR.

Figure~\ref{fig1} provides a visual illustration of harmonic and rational harmonic driving for selected ratios of $m$ and $n$. The red arrows indicate the input pump pulses, whilst the blue, green, and orange arrows correspond to excited solitons. Figure~\ref{fig1}(a) shows the conventional synchronous pumping scenario~\cite{Obrzud2017}, in which the periodicity of the pump pulse train is set close to the cavity round trip time (such that $f_\mathrm{in} \approx \text{FSR}$). Here, one cavity soliton circulates inside the resonator and is repumped by an external pulse every round trip, yielding an output comb with a frequency spacing of one FSR. Figure~\ref{fig1}(b) shows an example of harmonic driving with $f_\mathrm{in} \approx 3\,\text{FSR}$; the cavity contains three equally spaced solitons that are each repumped every round trip, yielding an output comb with a repetition rate three times the FSR. Figures~\ref{fig1}(c) and (d) show examples of rational harmonic driving. In Fig.~\ref{fig1}(c), we have $f_\mathrm{in} = (3/2)\text{FSR}$, yielding three equally spaced solitons which are repumped only every second round trip; in Fig.~\ref{fig1}(d), we have $f_\mathrm{in} = (2/3)\text{FSR}$, yielding two equally spaced solitons which are repumped every third round trip. In these last two cases [Figs.~\ref{fig1}(c) and (d)], the repetition rate of the input pulse train is three-halves and two-thirds of an FSR, yet yield output combs with a repetition rate close to three and two times the FSR, respectively. This illustrates the ability of rational harmonic driving to generate combs with multiple FSR frequency spacings, whilst driving at sub-FSR pump frequencies.

\begin{figure*}[t]	
		\includegraphics[width=\textwidth]{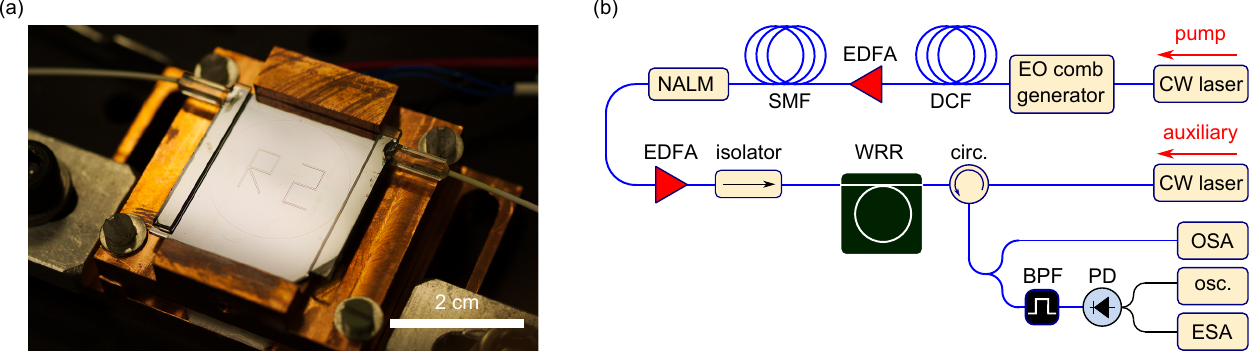}
		\caption{(a) Photograph of the waveguide ring resonator (WRR) used in our experiments. (b) Schematic illustration of the experimental setup. EO, electro-optic; DCF, dispersion compensating fiber; SMF, single-mode fiber; NALM, nonlinear amplifying loop mirror; EDFA, Erbium doped fiber amplifier; circ., circulator; OSA, optical spectrum analyzer; BPF, band-pass filter; PD, photodetector; ESA, electrical spectral analyzer; osc., oscilloscope.}
		\label{fig2}
\end{figure*}

Before proceeding to our experimental demonstrations, we comment on two salient technical details. First, when $n>1$, the solitons circulating in the cavity are not repumped every round trip, which leads to a reduction in the system's effective driving power~\cite{Coen2013_2}
\begin{equation}
X = \frac{\gamma P L\theta}{\alpha^3},
\end{equation}
where $P$, $\gamma$, and $\theta$ correspond to the peak power of the input pump pulse, the Kerr nonlinearity coefficient, and the input coupling coefficient, respectively. (Solitons can exist only for $X\gtrsim 2$~\cite{leo_dynamics_2013}, and their attainable spectral width scales as $\sqrt{X}$~\cite{Coen2013_2}.) Under conditions of rational harmonic driving, the parameters $L$ and $\alpha$ correspond, respectively, to the total length propagated and half of the total losses experienced by the intracavity field over one full cycle of repumping, i.e., $n$ round trips. It should be clear that $L = nL_0$ and $\alpha = n\alpha_0$, where $L_0$ and $\alpha_0$ are the corresponding (circumference) length and losses over one round trip, respectively, thus implying that the effective driving power $X\propto n^{-2}$. Accordingly, for a given resonator system (with fixed $\gamma$, $L_0$, $\alpha_0$, and $\theta$), the use of rational harmonic driving requires the pump peak power to be increased as $P\propto n^2$ so as to maintain constant \emph{effective} driving power $X$ (and hence potential comb bandwidth). Second, key to the harmonic and rational harmonic schemes is that all of the intracavity solitons are trapped at the same position with respect to their background pulse. Theories predict~\cite{hendry_spontaneous_2018}, however, that under conditions of perfect synchronisation [with respect to $(m/n)~\text{FSR}$], each pump pulse actually possesses \emph{two} stable trapping points (located on the pulse's leading and trailing edge respectively). Fortunately, this degeneracy can be lifted via appropriate desynchronization of the pump pulse train, forcing the system to favour only one of these two trapping points~\cite{Hendry2019}.

\section{Experimental Setup}

For experimental demonstration, we use an integrated waveguide ring resonator (WRR) formed from a buried Ge-doped low-loss silica-on-silicon waveguide [see Fig.~\ref{fig2}(a)]. Whilst such WRRs have been used in the past for gyroscope applications~\cite{Ma2017}, they have not (to the best of our knowledge) been applied for microcomb generation. The resonator is designed to support a single spatial mode around 1550~nm, and it has a diameter of 2~cm which corresponds to an FSR of 3.23~GHz. The Q-factor of the resonator is measured using the cavity ring-down method and found to be $2 \times 10^7$ at 1550~nm. Light is coupled to and from the resonator via an on-chip bus waveguide, which is itself coupled to standard single-mode optical fibers at each end of the bus. This provides for a very robust alignment-free resonator platform. More details on the fabrication of this device can be found in~\cite{Zhang2017,Lin2019}.

Figure~\ref{fig2}(b) shows our full experimental setup. We use an electro-optic (EO) frequency comb generator to create a picosecond pulsed laser source suitable for driving the WRR~\cite{Obrzud2017,anderson2019photonic}. Specifically, laser light from a narrow linewidth distributed-feedback fiber laser at 1550~nm is passed through one amplitude and two phase modulators driven by an RF signal generator. The resulting EO comb is then passed through 2.7~km of dispersion compensating fiber  to provide a stage of linear compression, followed by a nonlinear soliton compression stage consisting of an Erbium doped fiber amplifier (EDFA) and 1~km of single-mode fiber. The nonlinearly compressed pulse train is passed through a nonlinear amplified loop mirror to remove any unwanted pedestal from the pulse train~\cite{Agrawal_book}, then reamplified by a second EDFA to further increase the peak power. Frequency resolved optical gating is used to characterize the resulting pulses, which are found to have a full-width at half-maximum of 1.8~ps and a maximum peak power of 10 W. To provide easy access to the soliton state, an auxiliary CW laser at 1530 nm is used to provide thermal compensation~\cite{Zhang2019_2}. This auxiliary beam is launched into the WRR in the opposite direction to the pump, and it circulates in an orthogonally polarized resonator mode.

%With this setup, it is possible to directly tune the carrier frequency of the pulsed pump source into the effectively red-detuned regime that supports solitons.
\section{Results}

\begin{figure}[t]	
		\includegraphics[width=\linewidth]{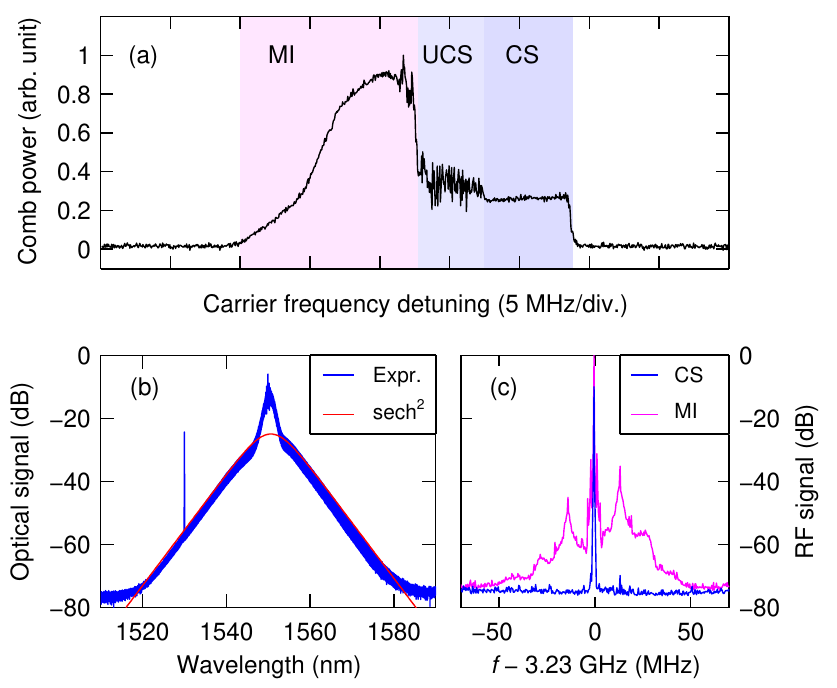}
		\caption{(a) Average comb power measured through an offset filter as the pump carrier frequency is scanned across a resonance. MI, modulation instability; UCS, unstable CS. (b) Measured comb spectrum when operating in the single soliton regime (blue curve) and $\text{sech}^2$ fit corresponding to a 300~fs CS (red curve). (c) RF beat note for a comb in the MI state (magenta) and CS state (blue curve). }
		\label{fig3}
\end{figure}

\begin{figure}[htb]	
		\includegraphics[width=\linewidth]{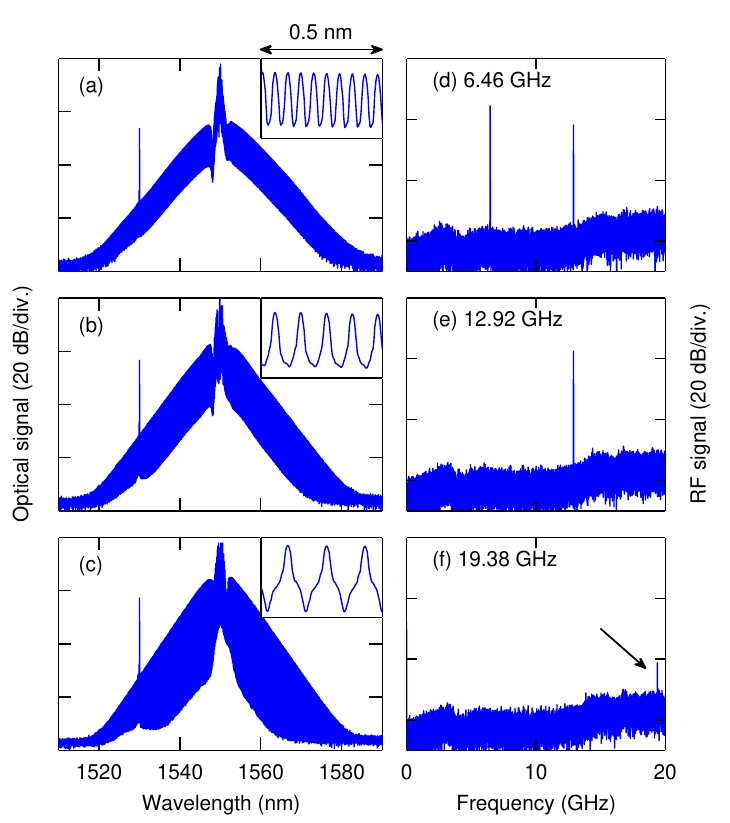}
		\caption{(a)--(c) optical spectra of the soliton comb when the resonator is pumped at 2, 4, and 6 times the cavity FSR, respectively. The insets show a zoom over a 0.5~nm spectral range centered around 1560~nm. (d)--(f) RF spectra corresponding to the optical spectra shown in (a)--(c), respectively.}
		\label{fig4}
\end{figure}

In order to first verify that our system can deterministically generate single soliton frequency combs, we set the pump's average power to 26 mW (corresponding to a peak power of about 4~W) and repetition rate close to the fundamental cavity FSR (3.23 GHz). Figure~\ref{fig3}(a) depicts a typical photodetector trace measured at the cavity output (and after a band-pass filter that excludes the pump) when the pump carrier frequency is scanned over a single resonance. We observe the usual signatures of stable CS formation: a region of modulation instability (MI) is followed by the formation of unstable (breathing) CSs and then a low-noise step indicating the presence of stable CSs. (Note that, throughout our experiments, the precise pump repetition rate is chosen so as to maximise the length of the soliton step, which also ensures that one of the intensity trapping positions is highly favoured due to the presence of stimulated Raman scattering~\cite{Hendry2019}.) Thanks to the auxiliary CW laser, the soliton step can be accessed in steady-state simply by manually advancing the cavity detuning. The output comb spectrum in this regime displays the $\text{sech}^2$ shape expected for a single soliton [see Fig.~\ref{fig3}(b)]. The small bump evident around 1550~nm is the remnant of the pump spectrum, whilst the sharp peak at 1530 nm comes from the backscattered component of the counterpropagating auxillary beam. To further confirm operation in the soliton regime, we used an electrical spectrum analyzer (ESA) to measure the fundamental RF beat note of the comb, observing a clear reduction in the RF noise upon entering the step region [see Fig.~\ref{fig3}(c)]. We must highlight that the resonator's FSR is sufficiently small such that fundamental RF beat signal can be readily resolved on a fast photodetector, allowing us to verify that the spacing of the soliton comb matches exactly with that of the electronic driving signal (to within the $\pm 10~\mathrm{Hz}$ accuracy of the ESA).

\begin{figure}[htbb]	
		\includegraphics[width=\linewidth]{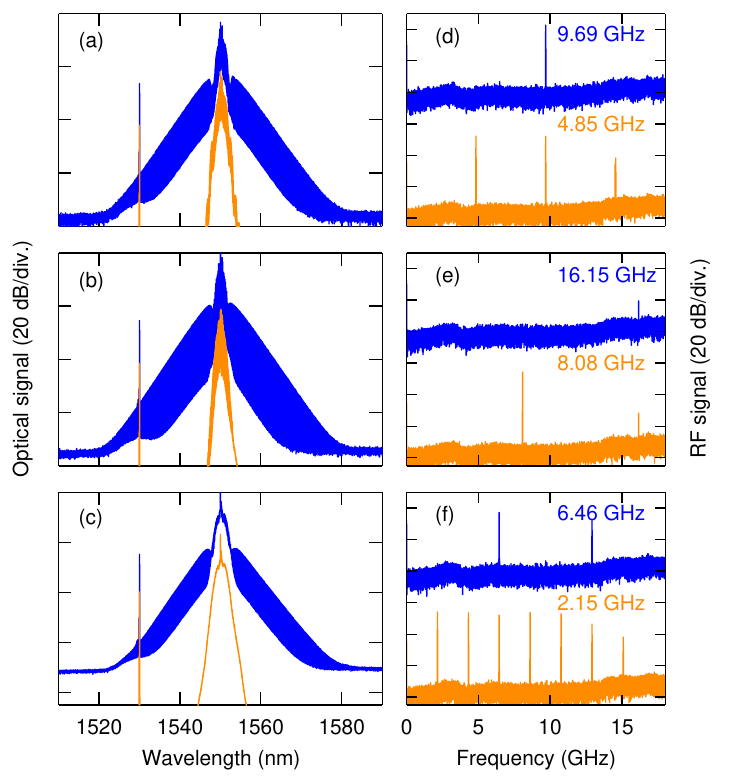}
		\caption{(a)--(c) Optical spectra of the soliton comb (blue curve) and the pump pulse train (orange curve) when the resonator is pumped at $3/2$, $5/2$, and $2/3$ of the cavity FSR, respectively. The corresponding electronic RF spectra are shown in (d)--(f). Note that the pump and soliton spectra are offset for clarity.}
		\label{fig5}
\end{figure}

To demonstrate harmonic driving, we set the pump repetition rate close to an integer multiple of the cavity FSR, and increase the average power so as to maintain peak power suitable for comb generation. For each of the repetition rates tested (up to six times the cavity FSR), we have identified and been able to manually tune into the low-noise soliton regime exactly as described above for the case of synchronous driving. Figures~\ref{fig4}(a)--(c) show illustrative examples of soliton comb spectra when the driving repetition rate was set to 2, 4, and 6 times the cavity FSR, respectively. Also shown are zoomed-in sections over a 0.5~nm spectral range to illustrate how the comb spacings adopt the pump repetition rates of 6.46~GHz, 12.92~GHz, and 19.38~GHz. For each case, we have also measured the RF spectrum of the output soliton comb [see Figs.~\ref{fig4}(d)--(f)], and only observe signals at integer multiples of the driving repetition rate. These measurements confirm that, thanks to the passive nature of the system, harmonic driving of soliton microresonator frequency combs is not associated with supermode noise that is often present in harmonically mode-locked lasers~\cite{Pottiez2002}.

To demonstrate rational harmonic driving, we set the pump repetition rate close to a rational fraction of the cavity FSR. As for synchronous and harmonic driving discussed above, we have observed the characteristic soliton step signature that can be manually accessed for all the rational fractions that we have tested. Figures~\ref{fig5}(a)--(c) show illustrative examples of soliton comb spectra when the pump repetition rate was set to 3/2, 5/2, and 2/3 of the cavity FSR, whilst Figs.~\ref{fig5}(d)--(f) show the corresponding RF spectra of the input pump (orange curve) as well as the output comb (blue curve). As can be seen, the results shown in Figs.~\ref{fig5}(a,d) and (b,e) demonstrate output combs with frequency spacings of 9.69 and 16.15~GHz obtained with a pump pulse train at 4.85~GHz (3/2~FSR) and 8.08~GHz (5/2~FSR), respectively. Note that the RF spectrum of the output comb is measured through a band-pass filter that excludes the pump, explaining why harmonics of the pump repetition rate that are not multiples of the cavity FSR do not appear in that spectrum. Finally, the results shown in Fig.~\ref{fig5}(c,f) remarkably demonstrate that driving the resonator with a repetition rate below the cavity FSR (here at $2.15~\mathrm{GHz}\approx 2/3\, \text{FSR}$) can allow for the generation of a soliton comb with a frequency spacing that is larger than the cavity FSR (here at $6.46~\mathrm{GHz}\approx 2\,\text{FSR}$) underlining the flexibility offered by rational harmonic driving.

Before closing, we note that both of the driving schemes demonstrated in our work are fully deterministic and highly repeatable. In particular, our experiments show that the target comb state is reached every time when manually tuning into the soliton state. This observation highlights the reliability of harmonic and rational harmonic driving in providing combs with desired line spacing.

\section{Conclusion and Discussion}

We have proposed and experimentally demonstrated harmonic and rational harmonic driving as flexible schemes to generate low-noise soliton microcombs with comb spacings that are close to integer multiples of the cavity FSR (and exactly locked to an external RF signal). Experimentally, using a single integrated silica waveguide ring resonator, we have demonstrated output combs with discretely adjustable frequency spacings between 3.23~GHz and 19.38 GHz. Moreover, we have shown that rational harmonic driving allows for the generation of output combs with multi-FSR frequency spacings whilst driving at sub-FSR frequencies. Measurements of the combs' RF spectrum confirm operation in the low-noise soliton regime with no supermode noise.

We believe that the new driving schemes demonstrated in our work could find useful application in areas, such as reconfigurable optical communication networks, where the ability to dynamically adjust the comb spacing of a transmitter or a receiver could prove beneficial. In addition, our results further underscore the fact that pulsed driving can enable soliton microcombs in resonator platforms that have not previously been considered suitable for that purpose, but that may nevertheless offer benefits in terms of ease-of-fabrication, absolute linewidth, or operational flexibility. As a forward-looking example, combining the novel driving schemes demonstrated in our work with high-finesse macroscopic optical fiber ring resonators could allow for the realization of novel sources of ultrashort pulses with GHz repetition rates discretely tunable in MHz steps.

\section*{Acknowledgements}
We acknowledge funding from the Marsden Fund, the Rutherford Discovery Fellowships, and the James Cook Fellowships of the Royal Society of New Zealand. We also acknowledge funding from the National Natural Science Foundation of China (61675181).

%\bibliography{WRR,rational}

\begin{thebibliography}{10}
\newcommand{\enquote}[1]{``#1''}

\bibitem{DelHaye2007}
P.~Del’Haye, A.~Schliesser, O.~Arcizet, T.~Wilken, R.~Holzwarth, and T.~J.
  Kippenberg, \enquote{Optical frequency comb generation from a monolithic
  microresonator,} Nature \textbf{450}, 1214--1217
  (2007).

\bibitem{kippenberg_microresonator-based_2011}
T.~J. Kippenberg, R.~Holzwarth, and S.~A. Diddams,
  \enquote{Microresonator-{Based} {Optical} {Frequency} {Combs},}
  Science \textbf{332}, 555--559 (2011).

\bibitem{pasquazi_micro-combs:_2018}
A.~Pasquazi, M.~Peccianti, L.~Razzari, D.~J. Moss, S.~Coen, M.~Erkintalo, Y.~K.
  Chembo, T.~Hansson, S.~Wabnitz, P.~Del{\textquoteright}Haye, X.~Xue, A.~M.
  Weiner, and R.~Morandotti, \enquote{Micro-combs: {A} novel generation of
  optical sources,} Physics Reports \textbf{729},
  1--81 (2018).

\bibitem{Kippenberg2018}
T.~J. Kippenberg, A.~L. Gaeta, M.~Lipson, and M.~L. Gorodetsky,
  \enquote{Dissipative kerr solitons in optical microresonators,}
  Science \textbf{361}, eaan8083 (2018).

\bibitem{gaeta_photonic-chip-based_2019}
A.~L. Gaeta, M.~Lipson, and T.~J. Kippenberg, \enquote{Photonic-chip-based
  frequency combs,} Nature Photonics \textbf{13},
  158--169 (2019).

\bibitem{suh_microresonator_2016}
M.-G. Suh, Q.-F. Yang, K.~Y. Yang, X.~Yi, and K.~J. Vahala,
  \enquote{Microresonator soliton dual-comb spectroscopy,}
  Science \textbf{354}, 600--603 (2016).

\bibitem{dutt_-chip_2018}
A.~Dutt, C.~Joshi, X.~Ji, J.~Cardenas, Y.~Okawachi, K.~Luke, A.~L. Gaeta, and
  M.~Lipson, \enquote{On-chip dual-comb source for spectroscopy,}
  Science Advances \textbf{4}, e1701858 (2018).

\bibitem{Lamb2018}
E.~S. Lamb, D.~R. Carlson, D.~D. Hickstein, J.~R. Stone, S.~A. Diddams, and
  S.~B. Papp, \enquote{Optical-frequency measurements with a kerr microcomb and
  photonic-chip supercontinuum,} Physical Review
  Applied \textbf{9}, 024030 (2018).

\bibitem{Spencer2018}
D.~T. Spencer, T.~Drake, T.~C. Briles, J.~Stone, L.~C. Sinclair, C.~Fredrick,
  Q.~Li, D.~Westly, B.~R. Ilic, A.~Bluestone, N.~Volet, T.~Komljenovic,
  L.~Chang, S.~H. Lee, D.~Y. Oh, M.-G. Suh, K.~Y. Yang, M.~H.~P. Pfeiffer,
  T.~J. Kippenberg, E.~Norberg, L.~Theogarajan, K.~Vahala, N.~R. Newbury,
  K.~Srinivasan, J.~E. Bowers, S.~A. Diddams, and S.~B. Papp, \enquote{An
  optical-frequency synthesizer using integrated photonics,}
  Nature \textbf{557}, 81--85 (2018).

\bibitem{bao_microresonator_2019}
C.~Bao, M.-G. Suh, and K.~Vahala, \enquote{Microresonator soliton dual-comb
  imaging,} Optica \textbf{6}, 1110--1116 (2019).

\bibitem{Trocha2018}
P.~Trocha, M.~Karpov, D.~Ganin, M.~H.~P. Pfeiffer, A.~Kordts, S.~Wolf,
  J.~Krockenberger, P.~Marin-Palomo, C.~Weimann, S.~Randel, W.~Freude, T.~J.
  Kippenberg, and C.~Koos, \enquote{Ultrafast optical ranging using
  microresonator soliton frequency combs,} Science
  \textbf{359}, 887 (2018).

\bibitem{suh_soliton_2018}
M.-G. Suh and K.~J. Vahala, \enquote{Soliton microcomb range measurement,}
  Science \textbf{359}, 884--887 (2018).

\bibitem{MarinPalomo2017}
P.~Marin-Palomo, J.~N. Kemal, M.~Karpov, A.~Kordts, J.~Pfeifle, M.~H.~P.
  Pfeiffer, P.~Trocha, S.~Wolf, V.~Brasch, M.~H. Anderson, R.~Rosenberger,
  K.~Vijayan, W.~Freude, T.~J. Kippenberg, and C.~Koos,
  \enquote{Microresonator-based solitons for massively parallel coherent
  optical communications,} Nature \textbf{546},
  274--279 (2017).

\bibitem{Herr2013}
T.~Herr, V.~Brasch, J.~D. Jost, C.~Y. Wang, N.~M. Kondratiev, M.~L. Gorodetsky,
  and T.~J. Kippenberg, \enquote{Temporal solitons in optical microresonators,}
  Nature Photonics \textbf{8}, 145 (2013).

\bibitem{leo_temporal_2010}
F.~Leo, S.~Coen, P.~Kockaert, S.-P. Gorza, P.~Emplit, and M.~Haelterman,
  \enquote{Temporal cavity solitons in one-dimensional {Kerr} media as bits in
  an all-optical buffer,} Nature Photonics \textbf{4},
  471--476 (2010).

\bibitem{Coen2013}
S.~Coen and M.~Erkintalo, \enquote{Universal scaling laws of kerr frequency
  combs,} Optics Letters \textbf{38}, 1790--1792
  (2013).

\bibitem{Coen2013_2}
S.~Coen, H.~G. Randle, T.~Sylvestre, and M.~Erkintalo, \enquote{Modeling of
  octave-spanning kerr frequency combs using a generalized mean-field
  lugiato-lefever model,} Optics Letters \textbf{38},
  37--39 (2013).

\bibitem{erkintalo_coherence_2014}
M.~Erkintalo and S.~Coen, \enquote{Coherence properties of {Kerr} frequency
  combs,} Optics Letters \textbf{39}, 283--286 (2014).

\bibitem{yi_soliton_2015}
X.~Yi, Q.-F. Yang, K.~Y. Yang, M.-G. Suh, and K.~Vahala, \enquote{Soliton
  frequency comb at microwave rates in a high-{Q} silica microresonator,}
  Optica \textbf{2}, 1078--1085 (2015).

\bibitem{brasch_photonic_2016}
V.~Brasch, M.~Geiselmann, T.~Herr, G.~Lihachev, M.~H.~P. Pfeiffer, M.~L.
  Gorodetsky, and T.~J. Kippenberg, \enquote{Photonic chip{\textendash}based
  optical frequency comb using soliton {Cherenkov} radiation,}
  Science \textbf{351}, 357--360 (2016).

\bibitem{webb_experimental_2016}
K.~E. Webb, M.~Erkintalo, S.~Coen, and S.~G. Murdoch, \enquote{Experimental
  observation of coherent cavity soliton frequency combs in silica
  microspheres,} Optics Letters \textbf{41},
  4613--4616 (2016).

\bibitem{jang_synchronization_2018}
J.~K. Jang, A.~Klenner, X.~Ji, Y.~Okawachi, M.~Lipson, and A.~L. Gaeta,
  \enquote{Synchronization of coupled optical microresonators,}
  Nature Photonics \textbf{12}, 688--693 (2018).

\bibitem{guo_universal_2017}
H.~Guo, M.~Karpov, E.~Lucas, A.~Kordts, M.~H.~P. Pfeiffer, V.~Brasch,
  G.~Lihachev, V.~E. Lobanov, M.~L. Gorodetsky, and T.~J. Kippenberg,
  \enquote{Universal dynamics and deterministic switching of dissipative {Kerr}
  solitons in optical microresonators,} Nature Physics
  \textbf{13}, 94--102 (2017).

\bibitem{cole_kerr-microresonator_2018}
D.~C. Cole, J.~R. Stone, M.~Erkintalo, K.~Y. Yang, X.~Yi, K.~J. Vahala, and
  S.~B. Papp, \enquote{Kerr-microresonator solitons from a chirped background,}
  Optica \textbf{5}, 1304--1310 (2018).

\bibitem{xue_microresonator_2017}
X.~Xue, P.-H. Wang, Y.~Xuan, M.~Qi, and A.~M. Weiner, \enquote{Microresonator
  {Kerr} frequency combs with high conversion efficiency,}
  Laser \& Photonics Reviews \textbf{11}, 1600276
  (2017).

\bibitem{Cole2017}
D.~C. Cole, E.~S. Lamb, P.~Del’Haye, S.~A. Diddams, and S.~B. Papp,
  \enquote{Soliton crystals in kerr resonators,} Nature
  Photonics \textbf{11}, 671--676 (2017).

\bibitem{Karpov2019}
M.~Karpov, M.~H.~P. Pfeiffer, H.~Guo, W.~Weng, J.~Liu, and T.~J. Kippenberg,
  \enquote{Dynamics of soliton crystals in optical microresonators,}
  Nature Physics \textbf{15}, 1071--1077 (2019).

\bibitem{Obrzud2017}
E.~Obrzud, S.~Lecomte, and T.~Herr, \enquote{Temporal solitons in
  microresonators driven by optical pulses,} Nature
  Photonics \textbf{11}, 600 (2017).

\bibitem{hendry_spontaneous_2018}
I.~Hendry, W.~Chen, Y.~Wang, B.~Garbin, J.~Javaloyes, G.-L. Oppo, S.~Coen,
  S.~G. Murdoch, and M.~Erkintalo, \enquote{Spontaneous symmetry breaking and
  trapping of temporal {Kerr} cavity solitons by pulsed or amplitude-modulated
  driving fields,} Physical Review A \textbf{97},
  053834 (2018).

\bibitem{brasch_nonlinear_2019}
V.~Brasch, E.~Obrzud, E.~Obrzud, S.~Lecomte, and T.~Herr, \enquote{Nonlinear
  filtering of an optical pulse train using dissipative {Kerr} solitons,}
  Optica \textbf{6}, 1386--1393 (2019).

\bibitem{obrzud_microphotonic_2019}
E.~Obrzud, M.~Rainer, A.~Harutyunyan, M.~H. Anderson, J.~Liu, M.~Geiselmann,
  B.~Chazelas, S.~Kundermann, S.~Lecomte, M.~Cecconi, A.~Ghedina, E.~Molinari,
  F.~Pepe, F.~Wildi, F.~Bouchy, T.~J. Kippenberg, and T.~Herr, \enquote{A
  microphotonic astrocomb,} Nature Photonics
  \textbf{13}, 31--35 (2019).

\bibitem{anderson2019photonic}
M.~H. Anderson, R.~Bouchand, J.~Liu, W.~Weng, E.~Obrzud, T.~Herr, and T.~J.
  Kippenberg, \enquote{Photonic chip-based resonant supercontinuum,} arxiv:1909.00022  (2019).

\bibitem{Onodera1993}
N.~Onodera, A.~J. Lowery, L.~Zhai, Z.~Ahmed, and R.~S. Tucker,
  \enquote{Frequency multiplication in actively mode‐locked semiconductor
  lasers,} Applied Physics Letters \textbf{62},
  1329--1331 (1993).

\bibitem{Jun1997}
W.~Jun~Shan, J.~Goldhar, and G.~L. Burdge, \enquote{Active harmonic modelocking
  of an erbium fiber laser with intracavity fabry-perot filters,}
  Journal of Lightwave Technology \textbf{15},
  1171--1180 (1997).

\bibitem{Chiming2000}
W.~Chiming and N.~K. Dutta, \enquote{High-repetition-rate optical pulse
  generation using a rational harmonic mode-locked fiber laser,}
  IEEE Journal of Quantum Electronics \textbf{36},
  145--150 (2000).

\bibitem{Yoshida1996}
E.~Yoshida and M.~Nakazawa, \enquote{80~200 ghz erbium doped fibre laser using
  a rational harmonic mode-locking technique,}
 Electronics Letters \textbf{32}, 1370--1372 (1996).

\bibitem{leo_dynamics_2013}
F.~Leo, L.~Gelens, P.~Emplit, M.~Haelterman, and S.~Coen, \enquote{Dynamics of
  one-dimensional {Kerr} cavity solitons,} Optics
  Express \textbf{21}, 9180--9191 (2013).

\bibitem{Hendry2019}
I.~Hendry, B.~Garbin, S.~G. Murdoch, S.~Coen, and M.~Erkintalo, \enquote{Impact
  of desynchronization and drift on soliton-based kerr frequency combs in the
  presence of pulsed driving fields,} Physical Review
  A \textbf{100}, 023829 (2019).

\bibitem{Ma2017}
H.~Ma, J.~Zhang, L.~Wang, and Z.~Jin, \enquote{Development and evaluation of
  optical passive resonant gyroscopes,} Journal of
  Lightwave Technology \textbf{35}, 3546--3554 (2017).

\bibitem{Zhang2017}
J.~Zhang, H.~Ma, H.~Li, and Z.~Jin, \enquote{Single-polarization
  fiber-pigtailed high-finesse silica waveguide ring resonator for a resonant
  micro-optic gyroscope,} Optics Letters \textbf{42},
  3658--3661 (2017).

\bibitem{Lin2019}
Y.~Lin, J.~Zhang, H.~Ma, and Z.~Jin, \enquote{Evaluation of polarization
  characteristics of the fiber-pigtailed waveguide-type ring resonator and
  implications for resonant micro-optic gyroscopes,}
 Journal of Lightwave Technology \textbf{37},
  2425--2434 (2019).

\bibitem{Agrawal_book}
G.~P. Agrawal, \emph{Nonlinear fiber optics}, Optics and photonics (Academic
  Press, San Diego, 2001), 3rd ed.

\bibitem{Zhang2019_2}
S.~Zhang, J.~M. Silver, L.~Del~Bino, F.~Copie, M.~T.~M. Woodley, G.~N.
  Ghalanos, A.~. Svela, N.~Moroney, and P.~Del’Haye,
  \enquote{Sub-milliwatt-level microresonator solitons with extended access
  range using an auxiliary laser,} Optica \textbf{6},
  206--212 (2019).

\bibitem{Pottiez2002}
O.~Pottiez, O.~Deparis, R.~Kiyan, M.~Haelterman, P.~Emplit, P.~Megret, and
  M.~Blondel, \enquote{Supermode noise of harmonically mode-locked erbium fiber
  lasers with composite cavity,} IEEE Journal of Quantum
  Electronics \textbf{38}, 252--259 (2002).

\end{thebibliography}
%
%\section*{References}

%\bibliographyfullrefs{WRR,rational}

\end{document}